\documentclass[12pt,english]{iopart}
\usepackage{graphicx}
\usepackage{amssymb}

\makeatletter
\usepackage{iopams}
\usepackage{setstack}

\usepackage{iopams}  
\makeatother

\usepackage{babel}

\begin{document}

\title{Geometric frustration in small colloidal clusters}


\author{Alex Malins$^{1}$, Stephen Williams$^{2}$, Jens Eggers$^{3}$,
Hajime Tanaka$^{4}$ and C. Patrick Royall$^{1}$ }

\address{$^{1}$ School of Chemistry, University of Bristol, Bristol, BS8
1TS, UK}

\address{$^{2}$ Research School of Chemistry, The Australian National University,
Canberra, ACT 0200, Australia}

\address{$^{3}$ School of Mathematics University of Bristol University Walk
Bristol BS8 1TW, UK}

\address{$^{4}$ Institute of Industrial Science, University of Tokyo, 4-6-1
Komaba, Meguro-ku, Tokyo 153-8505, Japan}

\ead{paddy.royall@bristol.ac.uk}

\address{Received May 20th, 2009}
\begin{abstract}
We study the structure of clusters in a model colloidal system with
competing interactions using Brownian dynamics simulations. A 
short-ranged attraction drives clustering, while a weak, long-ranged
repulsion is used to model electrostatic charging in experimental
systems. The former is treated with a short-ranged Morse attractive
interaction, the latter with a repulsive Yukawa interaction. We consider
the yield of clusters of specific structure as a function of the strength
of the interactions, for clusters with $m=3,4,5,6,7,10$ and $13$
colloids. At sufficient strengths of the attractive interaction (around
10 $k_{B}T$), the average bond lifetime approaches the simulation timescale
and the system becomes nonergodic. For small clusters $m\leqslant5$
where geometric frustration is not relevant, despite nonergodicity,
for sufficient strengths of the attractive interaction the yield of
clusters which maximise the number of bonds approaches 100\%. However
for $m=7$ and higher, in the nonergodic regime we find a lower yield
of these structures where we argue geometric frustration plays a significant 
role. $m=6$ is a special case, where two structures, of octahedral 
and $C_{2v}$ symmetry compete, with the latter being favoured by
entropic contributions in the ergodic regime and by kinetic trapping
in the nonergodic regime. 
We believe that our results should be valid 
as far as the one-component description of the interaction potential 
is valid. 
A system with competing electrostatic repulsions and van der Waals attractions 
may be such an example. 
However, in some cases, the one-component description of the 
interaction potential may not be appropriate. 

\end{abstract}


\section{Introduction}

Clusters are a distinct state of matter which exhibit different structural
ordering relative to bulk materials~\cite{baletto2005}. Of particular
relevance to, for example, many biological systems such as viruses,
is their tendency to exhibit five-fold symmetry such as icosahedra
and decahedra~\cite{wales}. Recently there has been a surge of interest
in clusters formed in colloidal systems~\cite{segre2001,manoharan2003,stradner2004,campbell2005,sedgwick2004,dibble2006,wilking2006,glotzer2007,wilber2007,zerrouki2008,anthony2008,akcora2009},
which is expected to lead to the development of `colloidal molecules'
\cite{manoharan2003,zhang2005diamond,glotzer2007,noya2007,anthony2008}.
These may in turn provide novel functionalised materials \cite{manoharan2003,zhang2005diamond,glotzer2007,wilber2007,noya2007,anthony2008}.

Part of the attraction of studying colloidal dispersions is that,
although in principle they are rather complex multicomponent systems,
the spatial and dynamic asymmetry between the colloidal particles
(10 nm-1 $\mu$m) and smaller molecular and ionic species has led to
schemes where the smaller components are formally integrated out 
\cite{likos2001}.
This leads to a one-component picture, where only the \emph{effective}
colloid-colloid interactions need be considered. 
This is usually a good approximation, however, there are a few exceptions. 
To describe the effective interactions between charged colloidal particles, 
the screening of the counter-ions is treated on a linear response level 
resulting in the well-known screened Coulomb pair potential proposed by 
Derjaguin, Landau, Verwey and Overbeek (DLVO). In general, however, due to 
non-linear counter-ion screening, the effective interactions involve 
many-body contributions \cite{Lowen}, which violates the additivity 
of the potential. Furthermore, in strongly driven systems hydrodynamic interactions 
(velocity fields) modify the electrostatic interaction significantly 
\cite{Kodama,Araki}. 
Here we assume the one-component description. 

The colloid behaviour
in the original complex system may then be faithfully reproduced by
appealing to liquid state theory~\cite{hansen} and computer simulation 
\cite{frenkel}.
Since the shape of the particles is typically spherical, and the effective
colloid-colloid interactions may be tuned, it is often possible to
use models of simple liquids to accurately describe colloidal dispersions.
For example this approach has made it possible to reproduce Bernal
spirals seen experimentally~\cite{campbell2005,sciortino2005bernal}
in a system with competing short-ranged attractions and long-ranged
repulsions, leading to the idea that colloidal gels can be stablised
by repulsive interactions~\cite{sciortino2004,kroy2004,zaccarelli2007}. 
In addition to their own fascinating behaviour such as novel 
diffusion~\cite{anthony2008}
colloidal clusters are also predicted from theory and simulation to
exhibit hierarchical ordering such as lamellae \cite{tarzia2006},
cluster crystals~\cite{mladek2006} and may also undergo dynamical
arrest to form cluster glasses~\cite{sciortino2004,kroy2004,fernandez2009}.

Since colloids may be directly imaged at the single-particle level,
one may consider local intra-cluster structure, along with cluster-cluster
correlations, a level of detail seldom accessible to atomic and molecular
systems, except in the low-temperature regime \cite{li2008}. Meanwhile,
the behaviour of colloidal clusters, for example the global energy
minimum structure, should exhibit similarities to that of clusters of
Noble gas atoms as both colloids and Noble gases can be well described
by spherically symmetric attractive interactions~\cite{doye1995,mossa2004}.
Some of us recently compared direct imaging of colloidal clusters
with expectations from theory~\cite{klix2009}.
In the experiments, a significant deviation from expectations was
found, in particular a maximum yield of only around 10\% in structures
expected to minimise the potential energy for small $4\leqslant m\leqslant6$
clusters. This is surprising, as at these small sizes, there is little
geometrical frustration that might inhibit access to the ground state,
as would occur for larger clusters.

It is the purpose of this work to establish those cluster structures
we expect in the case of model colloidal systems by applying tried
and tested model interactions. We shall consider weak electrostatic
repulsions and a short-ranged attraction for similar parameters to
the experimental system. Typically, the former stems from 
electrostatic charge,
the latter from the addition of non-adsorbing polymer 
or van der Waals attractions, and drives
clustering. Rather than seeking minimum energy 
states~\cite{doye1995,mossa2004}, instead we shall try to mimic experimental
approaches, as a way to predict what sort of yields of desired clusters
we find as a function of interaction strength. Experiments on colloidal
systems typically start from a randomised initial state. Unlike molecular
systems, the effective temperature is typically constant: temperature
itself is not usually varied and the effective colloid-colloid interactions
in a given sample are taken to be fixed, as they depend upon
sample composition. In other words the system undergoes an `instantaneous
quench' from an initial, randomised state. We follow this protocol,
although we note recent theoretical and simulation work highlighting
the role of microscopic reversibility in optimising yields in self-assembling
systems~\cite{whitesides2002,jack2007,rapaport2008} which is beginning
to be exploited in nanoscience \cite{nkypanchuk2008}. For the larger
cluster sizes considered, we also investigate the role of electrostatic
repulsions in cluster elongation, as suggested theoretically \cite{groenewold2001}
and noted experimentally \cite{stradner2004,campbell2005,sedgwick2004}.

Our approach is as follows. We shall consider clusters of $m=3,4,5,6,7,10,13$
particles and study the structures formed, with particular reference
to the minimum energy states ~\cite{doye1995,mossa2004}, as a function
of the attractive interaction strength (which mimics the addition
of polymer in experimental systems). Various strengths of the repulsive
interaction (which models the electrostatic charge in experimental
systems) are also considered. For high strengths of the attractive
interaction (low effective temperature), one expects the majority
of clusters to reside in their minimum energy state. However, average bond
lifetimes can exceed the simulation run time (which is comparable
to experimental timescales) for sufficiently strong interactions.
The system is thus nonergodic on these timescales, and kinetic trapping
may become important. For larger clusters, long-ranged repulsions
may lead to elongation \cite{stradner2004,campbell2005,sedgwick2004,sciortino2005bernal,mossa2004,groenewold2001}.
We investigate this effect for $m=10,13$. Our main result is that
for small clusters ($m\leqslant5$) the lack of geometric frustration
enables the minimum energy state to be accessed, while for larger
clusters kinetic trapping is important.

This paper is organised as follows: section \ref{section:model} introduces
the model interactions, our simulation methodology is presented in
section \ref{section:simulation}, followed by results (section \ref{section:results})
and a discussion in section \ref{section:discussion} in which we
place our findings in the context of recent work. We conclude our
findings in section \ref{section:conclusions}.

\section{Model}

\label{section:model}

\begin{figure}
\begin{centering}
\includegraphics[width=14cm]{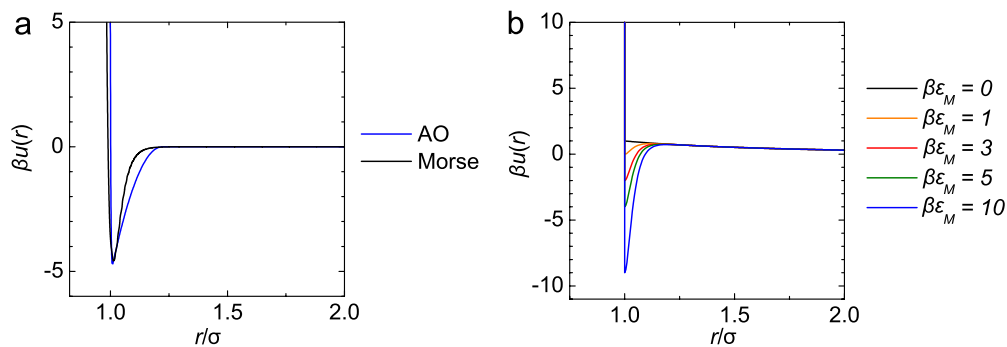} 
\par\end{centering}

\caption{(color online) (a) The Morse potential [eq. (\ref{eq:Morse}){]}
matched to the Asakura-Oosawa (AO) potential [eq. (\ref{eq:AO}){]}
relevant to experimental systems. (b) The competing interactions used
in this work, $\beta u_{M}+\beta u_{Y}$ for various values of the
Morse potential well depth $\beta\varepsilon_{M}$ as shown in the
legend. The case of $\beta\varepsilon_{M}=0$ is the pure Yukawa interaction
[eq. (\ref{eq:Yuk}){]} contact potential of $\beta\varepsilon_{Y}$=$kT$
and inverse Debye screening length $\kappa\sigma=0.5$.}

\label{fig:U} 
\end{figure}

The seminal theory of colloid-polymer mixtures is that of Asakura
and Oosawa~\cite{asakura1954}. This AO model ascribes an effective
pair interaction between two colloidal hard spheres in a solution
of ideal polymers which is plotted in figure \ref{fig:U} and reads,

\begin{equation} \beta u_{AO}(r)=\cases{\infty&for $r \le \sigma,$\\ \frac{\pi(2R_{G})^{3}z_{PR}}{6}\frac{(1+q)^{3}}{q^{3}}&\\ \times\{1-\frac{3r}{2(1+q)\sigma}+\frac{r^{3}}{2(1+q)^{3}\sigma^{3}}\}&for $\sigma< r\le\sigma+(2R_{G}),$\\ 0&for $r > \sigma+(2R_{G}),$\\} \label{eq:AO}\end{equation}


\noindent where $\beta=1/k_{B}T$, $r$ is the centre to centre separation of the two
colloids and the polymer fugacity $z_{PR}$
is equal to the number density $\rho_{PR}$ of ideal polymers in a
reservoir at the same chemical potential as the colloid-polymer mixture.
Thus within the AO model the effective temperature is inversely proportional
to the polymer reservoir concentration. The polymer-colloid size ratio
$q=2R_{G}/\sigma$ where $R_{G}$ is taken as the polymer radius of
gyration, and $\sigma$ is the colloid diameter. For small polymer-colloid
size ratios, the AO model has been found to give good agreement with
direct experimental measurement~\cite{royall2007}.

The discontinuous nature of the AO interaction at contact complicates
its use in Brownian dynamics simulations. We therefore use the continuous
Morse potential, which, for short interaction ranges, is very similar
to the Asakura-Oosawa potential (figure \ref{fig:U})~\cite{royall2008gel}.
The Morse potential reads

\begin{equation}
\beta u_{M}(r)=\beta \varepsilon_{M}e^{\rho_{0}(\sigma-r)}(e^{\rho_{0}(\sigma-r)}-2),\label{eq:Morse}\end{equation}

\noindent where $\rho_{0}$ is a range parameter and $\beta \varepsilon_{M}$ is the potential well depth. The global energy
minimum structures for clusters interacting via the Morse interaction
are known~\cite{doye1995}, and for small clusters $m<8$, the structure
of the global energy minimum is not sensitive to the range of the
interaction. With the exception of $m=6$ (see below), small ground
state clusters should be the same for an AO colloid-polymer system
as those tabulated for the Morse interaction. The experimental system~\cite{klix2009}
used a polymer-colloid size ratio of $q=0.22$ which maps to a Morse
range parameter $\rho_{0}=33.06$ for a well depth $\beta\varepsilon_{M}=5.0$
according to the extended law of corresponding states~\cite{noro2000}.

Repulsions in colloidal systems typically stem from the electrostatic
charge on the colloidal particles. Under many conditions, where the
charging is quite weak, as is the case here, this leads to a screened
Coulomb, or Yukawa form with a hard core that accounts for the physical
size of the colloidal particles:

\begin{equation} \beta u_{Y}(r)=\cases{\infty&for  $r < \sigma$,\\ \beta \varepsilon_{Y}\frac{\exp(-\kappa(r-\sigma))}{r/\sigma}&for $r\ge\sigma$,\\} \label{eq:Yuk}\end{equation}


\noindent where $\kappa$ is the inverse
Debye screening length. The contact potential is given by

\begin{equation}
\beta\varepsilon_{Y}=\frac{Z^{2}}{(1+\kappa\sigma/2)^{2}}\frac{l_{B}}{\sigma},\label{eq:BetaEpsilonYuk}\end{equation}

\noindent where $Z$ is the colloid charge and $l_{B}$ is the Bjerrum length. To model
the experimental system, we therefore fix the Debye length to an experimentally
relevant value $\kappa\sigma=0.5$, and consider different values
of the contact potential $\beta\varepsilon_{Y}$. In the experimental
system we seek to model, van der Waals attractions are largely absent,
due to solvent-colloid refractive index matching, the attractions are driven
by the addition of non-absorbing polymer.

We investigate the structure of these simulated colloidal clusters,
with reference to the Morse clusters~\cite{doye1995}, using a new
method we have developed, which we term the topological cluster classification
(TCC)~\cite{royall2008gel,williams2007}. Although we consider 
electrostatic repulsions, we expect little effect on the ground state for small $m<10$ clusters
~\cite{mossa2004}. 

\section{Simulation and analysis}

\label{section:simulation}

We use a standard Brownian dynamics simulation scheme~\cite{allen}.
The scheme generates a discrete coordinate trajectory ${\mathbf{r}_{i}}$
as follows\begin{equation}
\mathbf{r}_{i}(t+\delta t)=\mathbf{r}_{i}(t)+\frac{D}{k_{B}T}\sum_{j=1,j\neq i}^{N}\mathbf{F}_{ij}(t)\delta t+\delta\mathbf{r}_{i}^{G},\label{eq:BD}\end{equation}

\noindent where $\delta t$ is the simulation time step, $D$ is the diffusion constant and $\mathbf{F}_{ij}$
is the pairwise interaction. The colloids respond to
the direct interactions $\mathbf{F}_{ij}$ and the solvent-induced thermal
fluctuations $\delta\mathbf{r}_{i}^{G}$ 
are treated as a Gaussian noise with the variance
given by the fluctuation-dissipation theorem.

In the simulations $m$ particles are initialised randomly at volume
fraction $\phi=m(\pi\sigma^{3}/6)/l^{3}=0.0029$ in a cubic box of
side $l$.
Periodic boundary conditions for the box are implemented. We study
the evolution of one cluster per simulation, and consider the case
where all particle are part of the cluster. The inter-particle interaction
is the Morse potential [eq. (\ref{eq:Morse})] with range parameter
$\rho_{0}=33.06\sigma^{-1}$ which is truncated and shifted 
for $r>3\sigma$, 
where the Morse potential is typically less than $10^{-27}$. The
electrostatic interactions are treated by adding a Yukawa repulsion
term [eq. (\ref{eq:Yuk})] for different values of $\beta\varepsilon_{Y}$
as specified and this is also truncated and shifted for $r>3\sigma$.
This rather short value enables the particles to form clusters more
easily starting from the initial randomised state. While the Yukawa
repulsion has not fully decayed at $r=3\sigma$, the largest
separation of two particle centres we are interested in is set by
the size of the clusters. The largest separations for which we consider
the effects of the Yukawa repulsion are the $m=13$ clusters, where
the maximum separation is less than $3\sigma$. Since the Morse interaction
has no hard core, we remove the hard core component of eq. (\ref{eq:Yuk}). 
Each state point is sampled with between four and twelve statistically
independent simulation runs.

The time-step is $\delta t=0.03$ simulation time units and all runs
are equilibrated for $10^{9}$ steps and run for further $10^{9}$
steps. The rather long simulation runs were required to be sure that,
in the case of Yukawa repulsions, that the particles had ample time
to interact with one another, and cluster. We define the Brownian time 
as the time taken for a colloid to diffuse its own radius:

\begin{equation}
\tau_{B}=\frac{\sigma^{2}}{4D}.\label{eq:browntime}\end{equation}

\noindent In the simulations, $\tau_{B}\approx2474$ time units, while in the
experiments $\tau_{B}\sim10$ s~\cite{klix2009}. The simulation
runs therefore correspond to a total of around $68$ hours, a timescale
certainly comparable to experimental work.

We identify two particles as bonded if the separation of the particle
centres is less than $1.25\sigma$, which is close to the attractive
range of a strict AO potential {[}eq. (\ref{eq:AO}){]}. Having identified
the bond network, we use the Topological Cluster Classification (TCC)
to determine the nature of the cluster~\cite{williams2007}. This
analysis identifies all the shortest path three, four and five membered
rings in the bond network. We use the TCC to
find clusters which are global energy minima of the Morse potential.
We follow Doye \emph{et.al.} \cite{doye1995} and term these clusters
3A, 4A, 5A, 6A, 7A, 10B, and 13B for $m=3,4,5,6,7,10,13$. For $m\leqslant7$
there is one global minimum for the Morse potential. At higher $m$
there are multiple minima corresponding to different values of the
range parameter $\rho_{0}$. 10B and 13B correspond to a short ranged
Morse potential. We assume that these are the relevant global minima
for $\rho_{0}=33.06$. In addition, in the case of $m=13$ clusters
we identify the FCC and HCP thirteen particle structures in terms
of a central particle and its twelve nearest neighbours. For more
details see \cite{williams2007}.

\section{Results}

\label{section:results}

We shall consider each size in turn, before drawing together our results.
We use two control parameters, the well depth of the attractive Morse
interaction $\beta\varepsilon_{M}$ {[}eq. (\ref{eq:Morse}){]} and
the contact potential of the Yukawa repulsion $\beta\varepsilon_{Y}$
{[}eq. (\ref{eq:Yuk}){]}. Increasing $\beta\varepsilon_{M}$ thus promotes
clustering, while $\beta\varepsilon_{Y}$ suppresses clustering. In
experiments on colloids, the electrostatic charge is usually not systematically
varied. Therefore we consider specific values of $\beta\varepsilon_{Y}$
and plot the response of the system to $\beta\varepsilon_{M}$. Small
clusters with $m\leqslant5$ are able to reach the minimum energy
ground state structures (which maximise the number of bonds), while
for larger clusters, geometric frustration leads to kinetic trapping
which severely limits access to the ground state at high values of
the attractive interaction.

\subsection{Small clusters $m\leqslant5$}

\begin{figure}
\begin{centering}
\includegraphics[width=14cm]{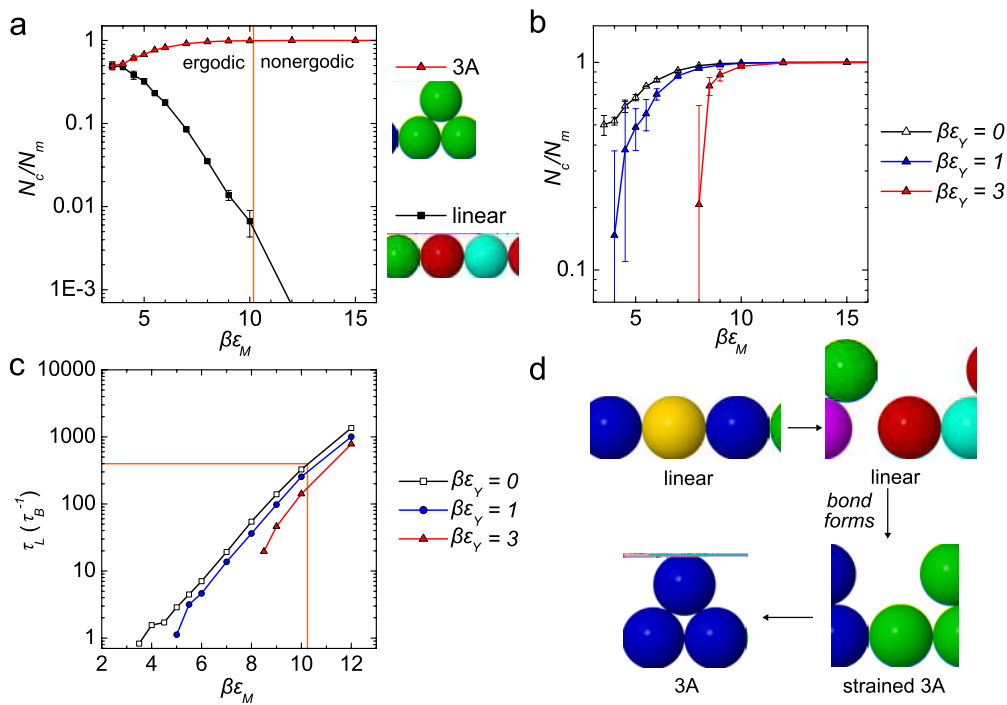} 
\par\end{centering}

\caption{(color online) (a) Cluster populations for the $m=3$ system as a
function of the well depth of the attractive interaction $\beta\varepsilon_{M}$
with $\beta\varepsilon_{Y}=0$. Here we define cluster population by the ratio
$N_{c}/N_{m}$ where $N_{c}$ and $N_{m}$ are the total number of specific clusters and 
the total number of $m$ membered clusters identified respectively in the simulation 
after equilibration. We see a clear trend towards
the 3A triangle as the minimum energy ground state structure. 
The orange line corresponds to $\tau_{L}=400\tau_{B}$,
our criterion for ergodicity breaking on the simulation timescale [see (c)].
(b) population of clusters in the minimum energy state (3A triangle)
for different strengths of the Yukawa interaction $\beta\varepsilon_{Y}$.
The principal effect of the Yukawa repulsion is to shift the curves
to require higher values of $\beta\varepsilon_{M}$ to achieve the same population
of clusters in the ground state. (c) the average bond lifetime $\tau_{L}$
as a function of $\beta\varepsilon_{M}$ for various $\beta\varepsilon_{Y}$
as indicated. The orange lines correspond to $\tau_{L}=400\tau_{B}$.
(d) the linear$\rightarrow$3A transition can be accomplished
by rotating one end particle around the central particle. This process
involves no energetic penalty for $\beta\varepsilon_{Y}=0$.}

\label{fig:m3} 
\end{figure}

We begin our presentation of results by considering the $m=3$ system,
as shown in figure \ref{fig:m3}(a). The main conclusion from these
data, as with all $m\leq5$, is that increasing the attractive interaction
strength $\beta\varepsilon_{M}$ leads to a higher population in the
minimum energy ground state, here the `3A' triangle with $D_{3h}$
point group symmetry and three bonds. This occurs at the expense
of higher energy clusters, which, for $m=3$ are `linear'
clusters with two bonds. In the case of $\beta\varepsilon_{Y}=0$,
the potential energy for a short-range potential such as eq. (\ref{eq:Morse})
is approximately equal to the number of bonds. Since 3A triangles
have three bonds, and the linear clusters have only two, there is a
strong thermodynamic driving force to the 3A cluster as the attractions
are increased, consistent with the behaviour seen in figure \ref{fig:m3}(a).

The effect of increasing the Yukawa repulsion is to slightly suppress
the development of the 3A population [figure \ref{fig:m3}(b)], as expected
in this system with competing interactions. In other words, the slight
upwards shift in the potential $\beta u_{M}+\beta u_{Y}$ {[}figure
\ref{fig:U}(b){]} due to the Yukawa contribution acts as a relatively
small perturbation to the $m=3$ system. However, 
$\beta\varepsilon_{Y}=5$
substantially suppressed the colloidal aggregation and few three-membered
clusters were formed. We hereafter consider $\beta\varepsilon_{Y}=0,1,3$
only.

The average bond lifetime $\tau_{L}$ is also shown in figure \ref{fig:m3}(c).
Here we define bond lifetime as the time between a bond formation
event (where the separation between two colloids falls below the bond
length $1.25\sigma$) a bond breakage event (where the separation
between the same two colloids rises above the bond length). The bond
lifetimes were widely distributed in all cases. We see that for $\beta\varepsilon_{M}\lesssim10$,
$\tau_{L}$ is very much less than the simulation time, so the system
may be regarded as equilibrated. In this regime, $\tau_{L}$ exhibits
an Arrhenius-like behaviour, as expected for an equilibrated system.
At higher values of the interaction strength, the average bond lifetime approaches,
and then exceeds the simulation run time, so the system is unable
to equilibrate, on the simulation (and experimental \cite{klix2009})
timescale. Thus the system is regarded as nonergodic on these timescales,
in that it cannot explore all its configurations. We take a average bond lifetime
$\tau_{l}=400\tau_{B}$ as a crossover between ergodic and nonergodic.
This ergodic - nonergodic transition is indicated in figure \ref{fig:m3}(a).
However, although the system is nonergodic, the absence of any geometric
frustration enables the minimum energy ground state to be reached.
The absence of geometric frustration is understood as follows {[}figure
\ref{fig:m3}(d){]}: for $m=3$, if $\beta\varepsilon_{Y}=0$ there
is no energy barrier in the linear$\rightarrow$3A transition, because
there is no angular dependence in the interaction. In other words,
a steepest descent quench for a three-membered cluster yields the
3A triangle.

\begin{figure}
\begin{centering}
\includegraphics[width=14cm]{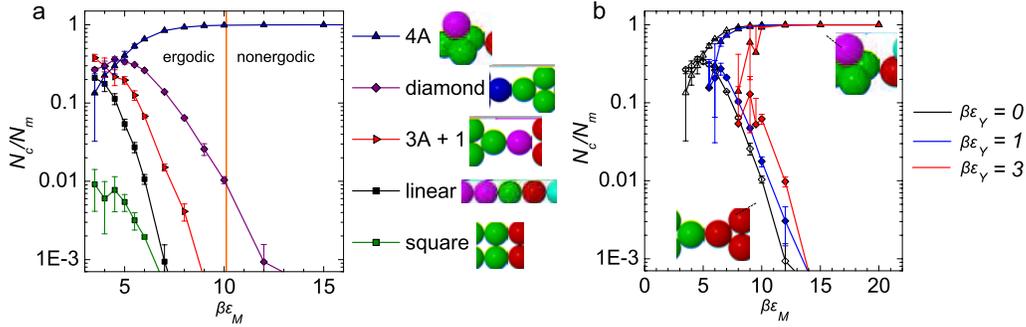} 
\par\end{centering}

\caption{(color online) (a) Cluster populations for the $m=4$ system as a
function of the well depth of the attractive interaction $\beta\varepsilon_{M}$
with $\beta\varepsilon_{Y}=0$. Here we see a clear trend towards
the 4A tetrahedron as the ground state structure. We identify three
higher energy structures: diamonds (5 bonds), triangle-lines (or 3A+1)
(4 bonds) and linear (3 bonds). The orange line corresponds to $\tau_{l}=400\tau_{B}$.
(b) population of clusters in the minimum energy ground state (4A
tetrahedron) and extended diamond structure for different strengths
of the Yukawa repulsion $\beta\varepsilon_{Y}$. The principle effect
of the Yukawa repulsion is to shift the curves to require higher interaction
strengths to achieve the same degree of clusters in the ground state:
the more elongated diamond structure does not appear more favoured
for stronger Yukawa repulsions.}

\label{fig:m4} 
\end{figure}

Turning to the case for $m=4$, the situation is somewhat more complex
{[}figure \ref{fig:m4}(a){]} Rather than two states, there are
four: 4A tetrahedra (the ground state with 6 bonds and $T_{h}$ point
group symmetry) diamonds (5 bonds), triangle-lines and squares (4
bonds) and linear (3 bonds). Squares are distinct from diamonds in
that there are no diagonal bonds. However, like the $m=3$ case,
increasing $\beta\varepsilon_{M}$ we find a peak in the population
of linear clusters, triangle-lines and diamonds respectively. Each
higher energy state has a distribution which becomes progressively
less favoured at higher values of $\beta\varepsilon_{M}$, as the
4A becomes the dominant structure, the yield of 4A approaches unity
for $\beta\varepsilon_{M}>10$. Squares have a rather low yield, much
less than triangle-lines (3A+1), which have the same number of bonds.
We return to the case of competing structures during our analysis
of $m=6$.

The effect of increasing the Yukawa interaction is similar to the
$m=3$ case: the development of the 4A population is somewhat suppressed
{[}figure \ref{fig:m4}(b){]}. It has been suggested that the introduction
of repulsions might be expected to promote more elongated structures
\cite{groenewold2001}. 
For a given strength of the attractive interactions $\beta\varepsilon_{M}$,
there is indeed a tendency towards a higher population of the elongated 4D diamond
structure, however the overall trend is unaltered and we restrict our analysis
to the $\beta\epsilon_{Y}=0$ case.

\begin{figure}
\begin{centering}
\includegraphics[width=10cm]{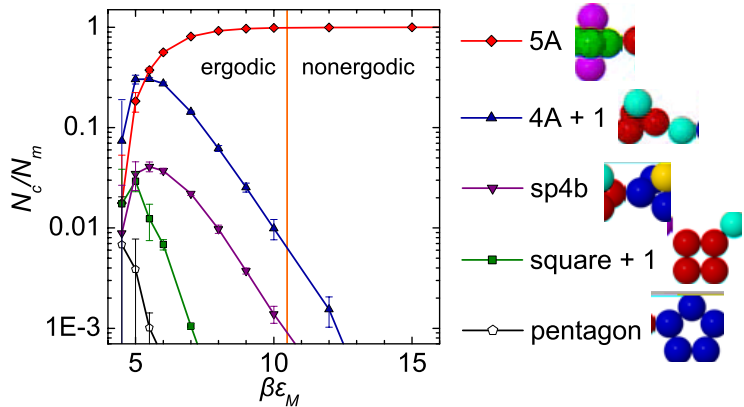} 
\par\end{centering}

\caption{(color online) Cluster populations for the $m=5$ system as a function
of the well depth of the attractive interaction $\beta\varepsilon_{M}$
with $\beta\varepsilon_{Y}=0$. The orange line corresponds to $\tau_{l}=400\tau_{B}$.
sp4b denotes a four-membered ring with one particle bonded to all
four in the ring.}

\label{fig:m5} 
\end{figure}

We now consider the $m=5$ system for $\beta\varepsilon_{Y}=0$ {[}figure
\ref{fig:m5}(a){]}. We find that for $\beta\varepsilon_{M}\gtrsim8.0$,
the overwhelming majority of clusters are in the ground state, 5A
(triangular bipyramid with 9 bonds and $D_{3h}$ point group symmetry),
in a similar way to the $m=3$ and $m=4$ cases. Defective triangular
bipyramids (4A+1) form an excited state with 10 or 11 bonds. Clusters
based around 4-membered rings with 5-8 bonds (sp4b and squares+1)
are present in yields up to a few percent for relatively weak attractions
($\beta\varepsilon_{M}\sim5$). Five-membered rings (pentagons, five
bonds) are found in small quantities, similar to the square in the
case of $m=4$. The main result from considering these small clusters
is that, although the system may become non-ergodic on the simulation
timescale, the clusters can
nevertheless access their global energy minima for sufficient interaction
strengths ($\beta\varepsilon_{M}\gtrsim8.0$). In other words, access
to the global minimum is not geometrically frustrated.

For $m=5$ we do not explicitly consider all possible structures. For example
linear clusters may form at weaker interaction strengths. At $\beta\varepsilon_{M}=4.5$,
for example, only around 10\% of $m=5$ clusters are identified, however
for most interaction strengths we consider, the vast majority of clusters
are one of the five structures considered. A similar argument holds
for $m=6$ and $m=7$.

\begin{figure}
\begin{centering}
\includegraphics[width=14cm]{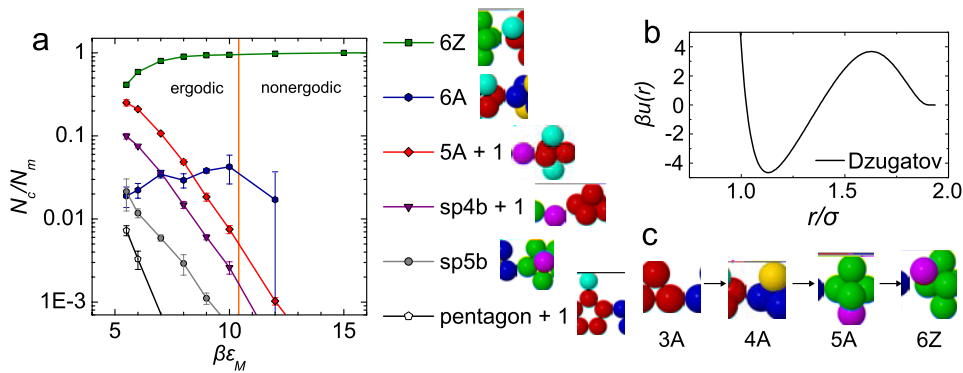} 
\par\end{centering}

\caption{(color online) (a) Cluster populations for the $m=6$ system as a
function of the well depth of the attractive interaction $\beta\varepsilon_{M}$
with $\beta\varepsilon_{Y}=0$. Here at strong interaction strength,
rather than the Morse ground state 6A octahedron, instead 6Z are found,
the ground state for the Dzugutov potential, (b). The orange
line corresponds to $\tau_{L}=400\tau_{B}$. (c) The 3A$\rightarrow$4A$\rightarrow$5A$\rightarrow$6Z
aggregation pathway with does not involve the breaking of any bonds
and thus promotes the formation of 6Z over 6A for high values of
$\beta\varepsilon_{M}$ where the average bond lifetime $\tau_{l}$} is comparable
to or greater than the simulation runtime.

\label{fig:m6} 
\end{figure}

\subsection{$m=6$ clusters: competing structures}

For $3\leq m\leq5$ the dominant structure at high interaction strength
is the global energy minimum for the Morse interaction. However, in the
case of $m=6$, we find this is not the case {[}figure \ref{fig:m6}(a){]}.
We see only a small population of the Morse global energy minimum,
the 6A octahedron ($O_{h}$ point group symmetry) with another
structure of $C_{2v}$ point group symmetry appearing to dominate the
system. This cluster is the ground state for the Dzugutov potential
{[}plotted in figure \ref{fig:m6}(b){]} \cite{doye2001}, and we term
it the 6Z. Why should this Dzugutov cluster be more popular than 6A
octahedra? Both clusters have 12 bonds (near-neighours), and for a
Morse potential with a relatively short range as we use here, the
energy contribution from second-nearest neighbours is a factor of
$5.625\times10^{-8}$ of the total energy for the 6A (it is essentially
zero for the 6Z). In fact, when the Yukawa repulsions are added, 6Z
becomes energetically favoured as it is more elongated. Let us however
consider the case for neutral particles where $\beta\varepsilon_{Y}=0$.

The difference between the two clusters' ground state potential energies is slight, 
but there are two remaining contributions to the system's free energy to be considered. 
Firstly the contribution due to the vibrational modes of the clusters must be accounted for, 
and secondly, so must the volume of phase space which is accessed upon translating and 
rotating the clusters through the system volume. Assuming that the vibrational modes may 
be approximated as harmonic (valid at low temperatures), the vibrational contribution to 
the free energies can be computed using a standard normal mode analysis. We have done 
this and found the contribution due to the vibrational modes to be completely negligible. 
The contribution due to translation will be the same for both clusters, so need not be considered. 
This leaves the contribution due to rotating the clusters. For the rotation we must consider 
both the point group symmetry and the cluster's radius of gyration. Comparing the point group 
symmetries, we can see  that there are $24$ different ways to reorientate the 6A cluster which are 
merely a permutation of indistinguishable particles, while for the 6Z there 
are only two ways. In computing the entropy we must count each permutation of indistinguishable 
particles only once. For this reason, in computing the entropy, we are able to rotate the 
6Z cluster through a greater portion of the available phase space. This results in an increase 
in the 6Z cluster's population relative to the 6A cluster's by a factor of $12$. One could think of the 6Z cluster's 
lower symmetry as a form of increased disorder.  We also consider the radius of gyration: 
$R_{G}^{2}=\frac{1}{N}\sum_{i=1}^{6}(\mathbf{r}_{i}-\mathbf{r}_{cm})^{2}$
where $\mathbf{r}_{i}$ are the particle coordinates and $\mathbf{r}_{cm}$
is the centre of mass. $R_{G}$ is larger by a factor of approximately $1.06$ times for the 6Z cluster than it is for the 6A. 
So upon rotating the 6Z cluster its particles traverse a greater portion of the available 
phase space than is the case for the 6A cluster. This increases the entropy of the 6Z cluster, 
relative to that of the 6A, further favouring it. It seems plausible that this could 
account for the relative differences in the populations, i.e., that the 6Z is thermodynamically 
favoured over the 6A by a factor of 30 as shown in figure \ref{fig:m6}(a).

As $\beta\epsilon_{M}$ becomes very large, we would ultimately expect
a trend towards a 6A dominated population due to the (small) difference in potential energy.
However, on these simulation
timescales, the average bond lifetime is too long to enable the transition
to a 6A dominated population, and, since 6Z can be the result of a
3A$\rightarrow$4A$\rightarrow$5A$\rightarrow$6Z aggregation sequence
{[}Fig \ref{fig:m6}(c){]}, and the formation of 6A requires bond
breakage, at strong interaction strengths, 6Z dominates for kinetic
reasons.

Like the smaller clusters, for $m=6$, we identify different structures
which become significant at lower values of $\beta\varepsilon_{M}$.
In decreasing stability, these are clusters with two tetrahedra {[}which
we denote as 5A+1 in figure \ref{fig:m6}(a), with either 10 or 11 bonds{]},
defective octahedra (denoted as sp4b+1, with between 9 and 11 bonds),
defective pentagonal bipyramids (denoted as sp5b with 10 bonds) and
clusters formed of a five-membered ring with one bound particle (pentagon+1,
between 6 and 9 bonds). 

\subsection{Larger clusters: geometric frustration}

\begin{figure}
\begin{centering}
\includegraphics[width=12cm]{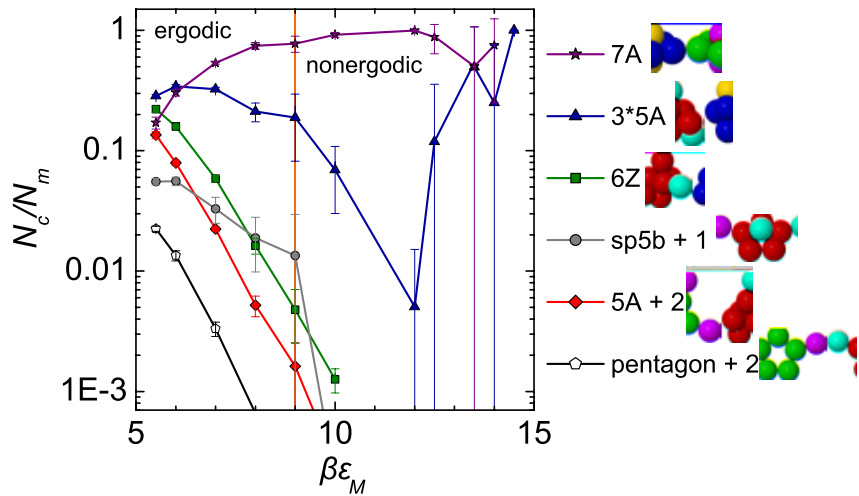} 
\par\end{centering}

\caption{(color online) Cluster populations for the $m=7$ system as a function
of the well depth of the attractive interaction $\beta\varepsilon_{M}$
with $\beta\varepsilon_{Y}=0$. Here the Morse ground state $7A$
pentagonal bipyramid is readily formed at intermediate interaction
strength. The orange line corresponds to $\tau_{l}=400\tau_{B}$.
sp5b denotes a five-membered ring with one particle bonded to all
five in the ring. A cluster based on three overlapping 5A triangular
bipyramids, 3$\times$5A, shows a re-entrant behaviour, dominating
the cluster population at high and low values of $\beta\varepsilon_{M}$.}

\label{fig:m7} 
\end{figure}

For the small clusters we have considered so far, the number of particles
is apparently too small to form metastable states. However, for $m=7$
we see that the yield of the Morse global energy minimum 7A pentagonal
bipyramid approaches unity for moderate strengths of $\beta\varepsilon_{M}$,
but for $\beta\varepsilon_{M}>12$ not all simulations reach the 7A
{[}figure \ref{fig:m7}(a){]}. Once the 7A is formed, the system remains
in a 7A state, but other metastable states have lifetimes longer than
the simulation runs. The rise in 7A population as a function of $\beta\varepsilon_{M}$
appears to slow around the ergodic-nonergodic transition (which we
define as $\tau_{L}=400\tau_{B}$). 

In the nonergodic regime where bond breaking governs those structures
which form, we expect an aggregation sequence similar to the 3A$\rightarrow$4A$\rightarrow$5A$\rightarrow$6Z
sequence shown in figure \ref{fig:m6}(c). Stepwise aggregation of one
particle onto a 6Z cluster would lead to a structure we term 3$\times$5A,
as it may be decomposed into 3 overlapping 5A triangular bipyramids,
or, equivalently, a 6Z with an additional
tetrahedron. This structure has 15 bonds. The 7A has 15 bonds where
the separation $\approx\sigma$, close to the minimum of the Morse
potential, while the distance between the top and bottom particle
is around $1.02283\sigma$, contributing $0.7193\varepsilon_{M}$
to the energy. The 7A is thus around $0.7193\beta\varepsilon_{M}$
lower in energy than the 3$\times$5A structure. In the nonergodic
regime for $\beta\varepsilon_{M}>12$, we see that this 3$\times$5A
structure dominates the system, due to kinetic trapping. The behaviour
of the 7A system in the nonergodic regime highlights the sampling
limitations of our simulation approach. To map this regime more accurately,
many more simulations are desirable than the $4$ runs per state point
we have been able to perform. 

At weaker interaction strengths, like the smaller clusters, a variety
of structures become popular, based on a diminishing number of bonds.
Of these, sp5b+1, a defective pentagonal bipyramid with 11-13 bonds
shows a rather slow rate of decay upon increasing $\beta\varepsilon_{M}$.
At the weakest interaction strength, $\beta\varepsilon_{M}=5.5$,
the yield of 3$\times$5A exceeds that of 7A. This `re-entrant' behaviour
of 3$\times$5A is apparently a consequence of the different number
of bonds relative to 7A. Cluster with fewer bonds are promoted for
weak interaction strengths, but in the nonergodic regime, kinetic
trapping promotes 3$\times$5A.

\begin{figure}
\begin{centering}
\includegraphics[width=14cm]{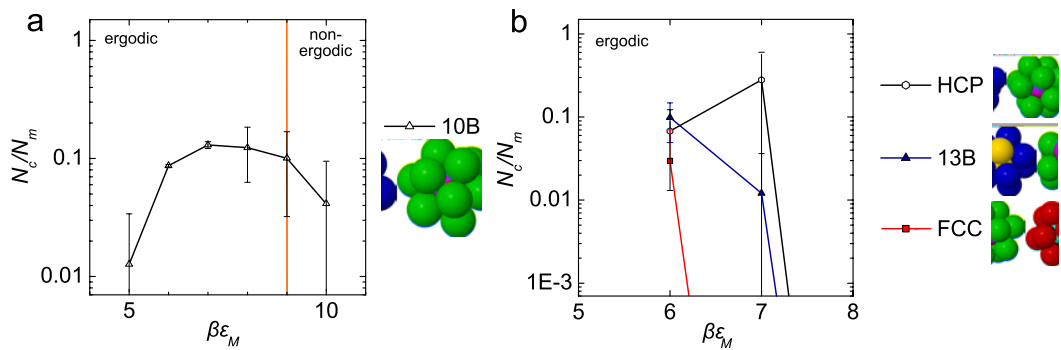} 
\par\end{centering}

\caption{(color online) (a) Cluster populations for the $m=10$ (a) and $m=13$
(b) system as a function of the well depth of the attractive interaction
$\beta\varepsilon_{M}$ with $\beta\varepsilon_{Y}=0$. The orange
line in (a) corresponds to\ $\tau_{l}=400\tau_{B}$}.

\label{fig:m10-13} 
\end{figure}

At larger sizes again we consider the Morse global minima, which are 10B for 13B 
for $m=10, 13$ clusters respectively. Once more the
effects of frustration are clear. The $m=10$ system (figure
\ref{fig:m10-13}) features a clear maximum yield of 10B at 
$\beta\varepsilon_{M}\sim7$, however the yield is only around 14\%.
The rise in the yield of 10B as a function of $\beta\varepsilon_{M}$
up to the ergodic-nonergodic crossover suggests that this rise in average
bond lifetime is the primary mechanism by which a further rise in
the 10B yield is suppressed. In other words, kinetic trapping prevents
access to 10B at higher strengths of the attractive interaction.

For $m=13$, in addition to 13B, we also find crystal fragments.
The maximum yield of 13B is around 10\%. For $\beta\varepsilon_{M}=7.0$,
HCP crystal fragments are in fact more popular than the 13B ground
state. We find no icosahedra, these are the global energy minimum
for longer ranged Morse interactions ($\rho_{0}<14.76$) \cite{doye1995}
than we use here, although for moderate values of $\beta\varepsilon_{M}$
it is not unreasonable to expect icosahedra. We expect that a similar
kinetic argument to that suggested above for $m=10$ concerning the
low yield of 10B holds for $m=13$ as well. For $m=10$ and $m=13$
we cannot exclude the possibility that other global minima exist.
Doye \emph{et. al.} \cite{doye1995} calculated global minima for
$\rho_{0}\lesssim25$. The smaller global energy minima for $m\leqslant6$
exhibit no strain for short-ranged Morse interactions, and only a
limited amount of strain in the case of 7A, these structures are therefore
also expected to be minima for $\rho_{0}=33.06$ as used here. This
not necessarily the case for $m=10$ and $m=13$. We are however unaware
of any more appropriate structures than those listed in \cite{doye1995}.

A question arises in comparing figure \ref{fig:m10-13} with results
for smaller clusters, such as the case of $m=3$ (figure \ref{fig:m3}).
In general one expects that larger clusters should be able to form
at higher temperatures (weaker interaction strengths) \cite{baletto2005},
(although for small Lennard-Jones clusters, the melting line is in
fact non-monotonic as a function of $m$ \cite{doye2002}). Here,
we are interested in whether all $m$ particles in the simulation
box aggregate to form a cluster. This is less likely for larger $m$,
so our statistics suffer for lower values of $\beta\varepsilon_{M}$
relative to the smaller clusters considered. It is nevertheless very
clear that the yield of the assumed global minimum clusters is markedly
reduced in the case of $m=10$ and $m=13$ relative to smaller clusters.
We now proceed to consider elongation.

\subsection{Elongation}

\begin{figure}
\begin{centering}
\includegraphics[width=14cm]{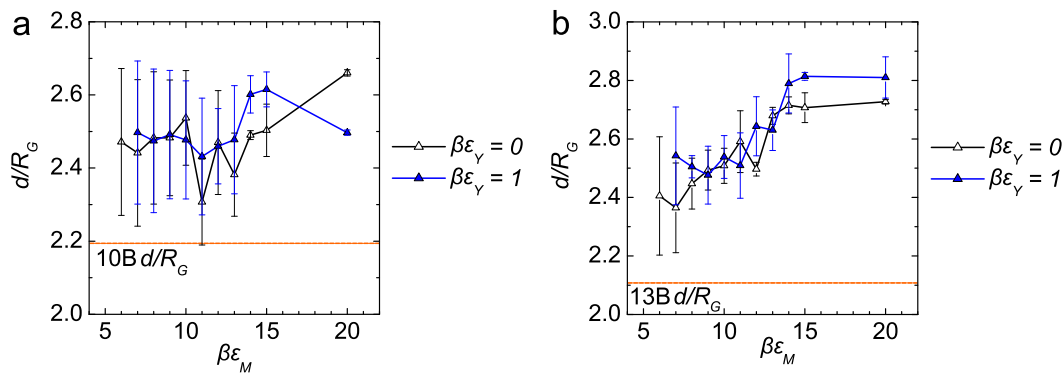} 
\par\end{centering}

\caption{(color online) (a) Elongation $d/R_{G}$ for the $m=10$ (a) and $m=13$
(b) system as a function of the well depth of the attractive interaction
$\beta\varepsilon_{M}$ with $\beta\varepsilon_{Y}=0$. The dashed
orange line denotes the elongation in the case of the minimum energy
ground state .}

\label{fig:m10-13-elongation} 
\end{figure}

The ground states of larger clusters have been found to be strongly
affected by the strength of a Yukawa repulsion, with stronger repulsions
leading to more elongated structures \cite{mossa2004}.
This effective long-ranged repulsion was indeed predicted to have a profound
effect upon both the shape and the size of clusters of charged colloids
by Groenewold and Kegel \cite{groenewold2001}, and experiments have
found evidence for elongated clusters \cite{stradner2004,sedgwick2004}
and Bernal spirals \cite{campbell2005}. We therefore investigate
the degree of elongation of the larger clusters considered here. We
consider the largest separation of two coordinates within a cluster,
$d$, and divide this by the radius of gyration $R_{G}$ of the same
coordinate set. Thus, larger $d/R_{G}$ corresponds to more elongated
clusters, and for a large spherical cluster with $m\rightarrow\infty$,
$d/R_{G}\rightarrow2\sqrt{5/3}$. For highly symmetric clusters such
as icosahedra, $d/R_{G}=2.08$.

Figure \ref{fig:m10-13-elongation} shows the elongation $d/R_{G}$
as a function of $\beta\varepsilon_{M}$. The elongation for $m=10$
and $m=13$ appears to be dominated by failure to access the ground
state; nonergodic systems are more elongated. The data for 
$\beta\varepsilon_{Y}=1.0$
are included in figure \ref{fig:m10-13-elongation}, while
for $\beta\varepsilon_{Y}=3.0$, few clusters formed
at all, and none in the ergodic regime. From the discussion above,
one might expect the $\beta\varepsilon_{Y}=1$ data to show a higher
degree of elongation due to the Yukawa repulsion, however 
figure \ref{fig:m10-13-elongation}(a)
indicates that this is not the case for $m=10$ and $m=13$, except
for a slight hint at the lower values of $\beta\varepsilon_{M}$.
We might expect that elongation induced by long-ranged repulsions
is promoted by increasing $m$, or $\beta\varepsilon_{Y}$ or perhaps
by a longer-ranged attraction that might suppress the transition to
non-ergodicity, due to a less complex potential energy surface (see
section \ref{section:discussion}) \cite{miller1999}. However in
this system we see little evidence for elongation.

\section{Discussion}

\label{section:discussion}

The behaviour of the systems considered here can readily be decomposed
into systems where frustration is not relevant, $m\leqslant5$ where
a well-defined structure is favoured, and those where frustration
leads to a non-trivial potential energy landscape and limited access
to the ground state $m\geqslant7$. In the case of $m=6$ two structures
with the same number of bonds, 6A octahedra and 6Z $D_{3h}$ compete.
Of particular relevance here is the short range of the attractive
interactions. It has been noted previously that these tend to promote
a complex energy landscape, for example $m=13$ Morse clusters with
$\rho_{0}=14.0$ have some $54439$ local energy minima, compared
to just $685$ for the longer-ranged case of $\rho_{0}=4.0$~\cite{miller1999}.
The Morse potential with $\rho_{0}=14.0$ is an approximation to $C_{60}$,
clusters of which are known to exhibit kinetic trapping \cite{baletto2002},
although in that case, icosahedral clusters were favoured. With $\rho_{0}=33.06$,
stronger trapping is likely. What is clear, therefore, is that these
systems can, only in the simplest cases ($m=3,4,5$), form high yields
of clusters in the minimum energy ground state, although $m=7$ does
have a window in $\beta\varepsilon_{M}$ where the yield of 7A is rather
substantial.

\subsection{Relevance to experiments}

This work has been largely motivated by experimental studies of clustering
in colloidal dispersions. In those systems, prized for their tunable
interactions, the attractions, either Asakura-Oosawa, or van der Waals
are typically rather short ranged, and thus likely to exhibit kinetic
trapping similar to the systems studied here. We therefore argue that
for substantial yields of more complex `colloidal molecules', it is
appropriate to seek more sophisticated means than the spherically
symmetric spheres we have considered here, such as patchy particles
\cite{glotzer2007,zerrouki2008,zhang2005diamond,nkypanchuk2008,park2008} and
Janus particles \cite{hong2006,hong2008},
or to design routes of preparation \cite{manoharan2003,sung2008}.
However, even `purpose-designed' patchy particles can have only a
rather limited window where the yield of ground state structures is
substantial~\cite{wilber2007,wilber2009,wilber2009b}.

We now compare our results to those found in an experimental system
in which the interactions are, to a first approximation, identical~\cite{klix2009}.
Each of the simulations was conducted with a fixed number of of particles
corresponding to the cluster size investigated. By contrast, most
experiments are carried out with a bulk colloidal suspension, in
which clusters of different size form. The experimental data
report the relative abundance of cluster types for a fixed number
of particles per cluster, $m$, and each cluster is assumed to interact
only weakly with other clusters. In the experimental system, $m=3$
formed a majority of clusters in the 3A triangle which maximises the
number of bonds. This occurred at interaction strengths comparable
to those found in the simulations reported here. In the experiments
$m=4$ and $5$ formed only around 10\% of 4A tetrahedra and 5A triangular
bipyramids respectively, in sharp contrast to the simulation results,
where the yield was essentially 100\%. In the case of $m=6$, both
in experiment and simulation, 6Z $C_{2v}$ dominated the 6A octahedron.
However, in the simulation, in the ergodic regime, the population
difference was around a factor of 30, while in the experiment the
difference was at least an order of magnitude larger. Furthermore,
the maximum yield of 6Z $C_{2v}$ was an order of magnitude lower
in the experiment than the simulation, where it was around 100\%.
For $m=7$ both experiment and simulation show signs of geometric
frustration, with the yield of 7A being reduced in the nonergodic
regime of strong interaction strength. However, while the peak yield
of 7A approached 100\% in the simulation, the experiments were limited
to a few percent of 7A. We consider possible origins for this discrepancy
in the following section.

\subsection{Charging in apolar colloidal systems}

Before concluding, it is worth considering these results in the light
of some other recent experimental studies. Campbell \emph{et. al.}~\cite{campbell2005}
report clusters in a system in which they measured the colloid charge
in a dilute suspension to be $Z=140$ $e$ per 1.5 $\mu$m diameter
colloid, where $e$ is the elementary charge. According to eq. (\ref{eq:Yuk}),
this maps to a Yukawa contact potential $\beta\varepsilon_{Y}=35$,
their quoted value for the Debye length is comparable to ours. We
have observed only very limited clustering at $\beta\varepsilon_{Y}=5$,
corresponding to a charge of $Z=47$ $e$ and expect none at higher
Yukawa contact potentials. Dibble \emph{et. al.}~\cite{dibble2006}
quote a similar value of $Z=165$ $e$ per 2 $\mu$m colloid. Moreover
Sedgwick \emph{et. al.}~\cite{sedgwick2004} report $Z<10^{3}$ $e$
in their study of clustering. Although not strictly inconsistent,
this seems rather higher than the values we predict from eq. (\ref{eq:Yuk}).

In their simulation study of gelation in colloidal systems with competing
interactions, Sciortino \emph{et. al.} \cite{sciortino2005} used
a comparable contact potential for the Yukawa repulsion $\beta\varepsilon_{Y}$
to that used in the simulations here. In other words, a much weaker
Yukawa repulsion than that expected from the colloid charge quoted
by Campbell \emph{et. al.} \cite{campbell2005}. Interestingly, Sciortino
\emph{et. al.} \cite{sciortino2005} found that similar Bernal spiral
structures were formed in their simulations to those observed
experimentally \cite{campbell2005}. From the arguments presented
above, it might be supposed that the Bernal spiral can form
without significant geometric frustration. An important question remains in
the role of colloid concentration. We have considered a rather low
volume fraction $\phi=m(\pi\sigma^{3}/6)/l^{3}=0.0029$, as we are
interested in isolated clusters. Whether the details of our results
apply at finite concentration where clusters interact with one another
remains to be seen, but many colloid volume fractions quoted in the
literature are $\phi\lesssim0.2$. In this regime we expect the discussion
above regarding clustering in the presence of electrostatic repulsions
to be reasonable.

Given the success of linear Poisson-Boltzmann theory in describing
electrostatic interactions in these systems in the absence of polymer-induced
attractions~\cite{royall2006}, one is tempted to enquire as to the
origin of the discrepancy between the findings of this paper and the
experimental literature~\cite{campbell2005,sedgwick2004,dibble2006}.
As far as we know, in only one case has quantitative agreement been
found between experiment and simulation of competing interactions
using conventional models of electrostatic repulsion and polymer-induced
attraction~\cite{royall2005}, and there the electrostatic repulsions
were screened by salt. Combined with the discrepancies presented here
and those in the experimental system~\cite{klix2009} to which we
have tuned the interactions, we believe that there is more than meets
the eye to these colloidal model systems with competing interactions.
We see little reason to suppose that the polymer-induced attractions
deviate substantially from theory \cite{royall2007}, therefore we speculate
that the electrostatic repulsions in the clusters may deviate from those
deduced by electrophoretic mobility measurements in dilute suspensions \cite{campbell2005,dibble2006,
klix2009,royall2006}.

\section{Conclusions}

\label{section:conclusions}

We have studied isolated colloidal clusters using Brownian dynamics simulations.
Our system is tuned to match experimental work in colloidal systems
with polymer-induced depletion attractions. For sufficient strengths
of the attractive interaction, the average bond lifetime exceeds the simulation
runtime, and the system as nonergodic on this timescale;
this conclusion also applies to some experimental systems. However,
for small clusters $m\leqslant5$ this ergodicity breaking does not
prevent the system reaching the minimum energy ground state structure,
as no bonds need be broken. Thus for these small clusters, we can
direct the system in a controlled way towards a prescribed ground
state. For $m=6$ we find a structure which appears to minimise the
free energy with $C_{2v}$ symmetry, rather than the octahedron that
forms the minimum energy ground state~\cite{doye1995,mossa2004}.
The energy difference between these structures is negligible, and
for moderate interaction strengths (the ergodic regime) the $C_{2v}$
structure is favoured for entropic reasons, while in the nonergodic
regime, it is kinetically favoured as it is the product of an aggregation
sequence that does not involve bond breakage. At larger cluster sizes,
the system is kinetically frustrated from reaching the ground state
in all but a limited window. Moreover, recent experimental results
in a similar system show much lower yields of clusters in the ground
state than we find here \cite{klix2009}. The origin of the discrepancy
is likely related to a difference in the effective interaction potentials
between experiment and simulation. It is perhaps surprising that such
small systems exhibit this degree of kinetic trapping; only those
sizes for which there is no geometric frustration ($m=3,4,5$) reach
the minimum energy ground state. These results suggest that high yields
of complex microstructures of colloidal clusters and molecules may
benefit from reversible quenching~\cite{whitesides2002,jack2007,rapaport2008,nkypanchuk2008},
rather than the fixed interactions of many colloidal systems, which
lead to `instantaneous quenching'.

Some pointers for further work are considered. We have employed a
simple approach in our simulations, as we are motivated to reproduce
the recent experimental system. One approach to investigate larger
clusters could be to run a simulation of a much larger number of particles,
and to consider each cluster as a separate system. In this way, larger
clusters could be accessed than was the case here. Although this could
in principle resolve some discrepancies between the results presented
here and those reported for the experimental system, we believe
this is a most unlikely scenario. Other possibilities might be to
implement the methods found in the (atomic) cluster literature \cite{baletto2005,doye1995,doye2002}
to comprehensively determine the structure of larger clusters. 
Another possibility would be to determine the phase diagram for the
system considered here, using normal mode analysis to provide a
theoretical prediction of the population levels for the various
structures considered.

To provide further support to experimentalists in the quest to control
the structure of colloidal clusters it might be helpful to move beyond
the one-component description employed here. Given that the charging
number quoted in some experimental work is so small ($Z\lesssim100$
$e$)\cite{campbell2005,sedgwick2004,dibble2006,klix2009,roberts2007},
it might be possible to determine structures of colloidal clusters
by for example developing the primitive model of electrolyte systems
to colloidal systems with many ionisable sites on each particle \cite{lobaskin1999}.

\section*{Acknowledgments}

This study was partly supported by EPSRC through grant code EP/E501214/1,
and by a grant-in-aid from the Ministry of Education, Culture, Sports,
Science and Technology, Japan. CPR acknowledges the Royal Society
for financial support.

\section*{References}


\end{document}